\documentclass[jog]{igs}

\usepackage[utf8]{inputenc}

\usepackage{graphicx}
\usepackage{siunitx}
\usepackage{xcolor}

\jourpubyear{2022}

\begin{document}

\title[Direct measurement of optical properties of glacier ice using a photon-counting diffuse LiDAR]{Direct measurement of optical properties of glacier ice using a photon-counting diffuse LiDAR}

\author[Allgaier and others]{Markus ALLGAIER$^1$, Matthew G. COOPER$^2$, Anders E. CARLSON$^3$, Sarah W. COOLEY$^4$, Jonathan C. RYAN$^4$, Brian J. SMITH$^1$}

\affiliation{%
$^1$Department of Physics and Oregon Center for Optical, Molecular, and Quantum Science, University of
Oregon, Eugene, OR 97403, USA\\
$^2$Pacific Northwest National Laboratory, Richland, WA 99354, USA\\
$^3$Oregon Glacier Institute, Corvallis, OR 97330, USA\\
$^4$Department of Geography, University of
Oregon, Eugene, OR 97403, USA\\

  Correspondence: Markus Allgaier
  \email{markusa@uoregon.edu}}

\begin{frontmatter}

\maketitle

\begin{abstract}
The production of meltwater from glacier ice, which is exposed at the margins of land ice during the summer, is responsible for a large proportion of glacier mass loss. The rate of meltwater production from glacier ice is especially sensitive to its physical structure and chemical composition which combine to determine the albedo of glacier ice. However, the optical properties of near-surface glacier ice are not well known since most prior work has focused on ice made in the laboratory or from deep cores. Here, we demonstrate a measurement technique based on diffuse propagation of nanosecond-duration laser pulses in near-surface glacier ice that enables the independent measurement of the scattering and absorption coefficients, allowing for a complete description of the processes governing radiative transfer. We employ a photon-counting detector to overcome the high losses associated with diffuse optics. The instrument is highly portable and rugged, making it optimally suited for deployment in remote regions. A set of measurements taken on Collier Glacier, Oregon, serves as a demonstration of the technique. These measurements provide insight into both physical structure and composition of near-surface glacier ice and open new avenues for the analysis of light-absorbing impurities and remote sensing of the cryosphere.
\end{abstract}

\end{frontmatter}

\section{Introduction}

Mountain glaciers, ice caps, and ice sheets have been shrinking since the late 20\textsuperscript{th}
and early 21\textsuperscript{st} century due to anthropogenic climate change and have contributed substantially to global sea-level rise \citep{meredith2019}. Much of this glacier mass loss is due to warmer summer air temperatures, which initiate positive feedbacks between albedo and melt that further enhance mass loss. For example, warmer air temperatures increase snow grain sizes, causing more absorption of shortwave radiation and further melt \citep{gardner2010review}. Other processes, particularly the deposition of light-absorbing particles (LAP) such as black carbon (BC) and dust are also known to reduce ice and snow albedo and enhance melt \citep{dumont2014,skiles2018}. More recently biological radiative forcing by algae has received much attention as a driver for enhanced snow and ice melt \citep{dial2018,Williamson2018,ryan2018dark}. Precise knowledge of how glacier ice optical properties respond to warming air temperatures, BC, dust, and algae is therefore a prerequisite for accurately modeling glacier and ice sheet melt and subsequent contribution to sea levels.
 
Understanding the optical properties of glacier ice is not only important for modeling the radiative transfer and energy budgets of ice and snow, but also for remote sensing. Scattering of light within snow and glacier ice, for example, may introduce a bias in laser altimetry \citep{smith2018,greeley2019}. With the surface reflection from ice accounting for only about 2\% of the total intensity, the majority of the backreflection must occur below the surface. Photons that scatter multiple times below the surface take longer to return to the sensor, making the surface appear lower than it actually is. This process is especially important for the ATLAS instrument onboard ICESat-2, which operates at 532 nm, a wavelength known to have low absorption and strong forward scattering in snow and ice \citep{warren2008}. If not properly corrected, this process could introduce a centimeter to decimeter bias in sea ice, glacier and ice sheet elevations derived from ICESat-2.

LAPs are of particular importance for albedo and subsequent melt rates of snow \citep{hadley2012,kaspari2015,zhang2017}. There is a significant body of work dedicated to describing absorption, scattering, crystal structure, impurities and apparent optical properties of snow \citep{warren1980a,warren1980b,kokhanovsky2004,libois2013,saito2019,beres2020}. These models accurately reproduce field measurements of snow albedo. Similarly, sophisticated models exist for sea ice \citep{light2010}. In contrast, relatively few studies have investigated the optical properties of glacier ice.  \cite{perovich1991} derived absorption coefficients for pure bubble-free ice between 250–600 nm using a laboratory transmission measurements. Likewise, \cite{ackermann2006} derived scattering and absorption coefficients of glacier ice by measuring the frequency distribution of photon travel times from a pulsed light source in glacier ice deep in the Antarctic Ice Sheet as part of the Antarctic Muon and Neutrino Detector Array (AMANDA) experiment. However, the clean, bubble-free ice investigated by these studies is likely not representative of glacier ice exposed in the ablation zones of most ice masses.
Reviews of pure ice properties derived from snow have been published by \cite{warren2019} and \cite{kokhanovsky2021}.

So far, in-situ measurements of scattering and absorption coefficients of glacier ice have remained challenging. Transmission measurements rely on assumptions about the wavelength-dependence of the scattering coefficient at visible wavelengths as well as the qualitative influence of LAPs, for which reference values are needed \citep{warren2006,cooper2020}. The same is true for scattering intensity measurements, where merely an effective constant describing both scattering and absorption can be retrieved \citep{allgaier2021}. Although an attempt was made to retrieve independent values for scattering and absorption from experimentally measured scattering intensity distributions using numerical simulations \citep{trodahl1987}, a truly independent measurement has only been shown in the framework of detector calibration for AMANDA and IceCube \citep{ackermann2006,aartsen2013}, where the propagation of short optical pulses through deep glacier ice at the South Pole was measured in the time domain. For this purpose, a diffusion transport model was developed \citep{askebjer1997}. However, this model was found to be inaccurate in bubble-free, deep ice, where the conditions to satisfy the diffusion approximation are not met due to insufficient scattering. Again, the solution was to use simulated photon propagation instead of the analytical model. Similar measurements performed on sea ice also cannot be described by the diffusion model since the detector is in the diffusion-near-field \citep{perron2021}, close to the source. Because the diffusion model has only two free parameters - effective scattering and absorption coefficients - it would be a convenient tool for measuring these properties on the surface of glaciers, sea ice and snow, where it would provide an independent measurement free of assumptions. Along with the refractive index, these two parameters  are sufficient to calculate the albedo of glacier ice \citep{dombrovsky2020}. When albedo is inferred in this way, it can be interpreted in terms of scattering and absorption, which is not possible from albedo measurements alone. Numerical simulations have shown that the diffusion model is a robust tool to accomplish these measurements on the surface if the appropriate boundary conditions are considered \citep{allgaier2021}.

Here, we present a new instrument that implements a time-resolved scattering experiment with a narrow-band point source on the surface. A photon-counting detector is placed in the scattering-far-field, several scattering lengths away from the source, which allows us to employ a diffusion model that includes appropriate boundary conditions to describe behavior at the surface, outlined in our recent theoretical analysis \citep{allgaier2021}. The device is highly portable, making it ideal for deployment in rugged and remote areas. We show data measured in the field on Crook and Collier Glacier in the Central Oregon Cascades in September 2021. We conclude with a discussion about how data obtained with this instrument could inform surface energy balance models with accurate scattering and absorption coefficients, provide LiDAR and laser altimeter penetration depth, and provide a proxy for measuring BC concentration in glacier ice.

First, we provide a summary of the employed diffusion model and its implications on experimental resolution as well as context about the established terminology for scattering in glacier ice and snow. Next, we describe the instrument and its performance in terms of resolution and range. We present the experimental results of the measurements on Crook and Collier Glacier. Finally, we discuss statistical correlations of the data, compare extracted coefficients to the literature, and showcase how the measured scattering and absorption coefficients can be used to infer spectral and broadband albedo as well as LAP concentration.

\section{Diffuse propagation of a point source}\label{sec:diffusion}

Diffuse imaging has been employed in tissue for over two decades, where it is known that retrieval of both scattering and absorption parameters requires a modulated or non-stationary source in either spectrum or time \citep{durduran2010}. Transport of light in scattering media, particularly glacier ice, takes place on fairly long time scales. Therefore, implementation of diffuse measurements in the time-domain is convenient. Here, we restrict ourselves to geometries where source and detector are far apart. In the presence of both scattering and absorption, radiative transport in terms of a fluence rate \(\phi(r,t)\), where \(r\) is the distance from the source and \(t\) is the time since injection of the instantaneous source, is described by the diffusion equation

\begin{equation}
    \frac{\partial \phi(r,t)}{\partial t} = D \nabla^2 \partial \phi(r,t) - c\beta\phi(r,t) + cS(r,t),
\end{equation}

\noindent in which \(c=c_0/n\) is the speed of light in the medium, \(c_0\) is the vacuum speed of light, \(n\) is the refractive index and \(S(R,T)\) is the source distribution. \(\beta=c/l_{\mathrm{abs}}=c\sigma_{\mathrm{abs}}\) accounts for absorption, and \(\sigma_{\mathrm{abs}}\) defines the medium's absorption coefficient. The fluence rate describes energy per time transported through a unit area of the medium. Integrated over a detector area, the fluence rate becomes power. The source function \(S(R,T)\) describes the extent of the light source in space and time (spatial mode shape and temporal pulse shape). Without loss of generality, we can assume the source to be infinitesimal in time and space, i.e. \(S(R,T)=\delta (T) \delta (R)\). For any extended source or finite pulse length, the following solutions for the infinitesimal source can simply be convoluted with the source function \cite{fantini1997}. \(D\) is the diffusion constant, which is related to the effective, isotropic scattering constant:

\begin{equation}
    D = \frac{c l_{\mathrm{eff}}}{3} = \frac{c }{3\sigma_{\mathrm{eff}}}.
\end{equation}

Here, \(l_{\mathrm{eff}}\) is the mean-free path of an effective, isotropic scattering process and the inverse scattering constant \(\sigma_{\mathrm{eff}}=1/l_{\mathrm{eff}}\). The well-known Green's function that solves this diffusion equation for an instantaneous point source at \(r=0,t=0\) inside an infinite medium is

\begin{equation}
\phi_{+}(r,t) = \frac{1}{\left(4\pi Dt\right)^{3/2}}\mathrm{exp} \left(-\frac{r^2}{4Dt}  - \beta t\right),
\label{eq:diffsolution}
\end{equation}

\noindent where 
\(\beta=c/l_{\mathrm{abs}}=c\sigma_{\mathrm{abs}}\) dampens the flux at long times through absorption, described by the absorption length \(l_{\mathrm{abs}}\) or absorption coefficient \(\sigma_{\mathrm{abs}}\). A detailed derivation of this solution and its relationship to a random walk in the photon picture can be found in \cite{askebjer1997}.

It is important to note that this description is only valid in the scattering-far-field, where \(r\gg l_{\mathrm{eff}}\), which practically means that a detector needs to be placed sufficiently far away from the source. To account for the fact that a realistic laser source is not isotropic, we replace it with an effective isotropic source that lies one scattering length below the surface. In reality, glacier ice is a compact, finite medium with refractive index \(n=1.31\) \citep{warren2019}, which means that a significant portion of the light impinging onto the surface from below is reflected back into the medium. To account for this, we adopt an appropriate boundary condition. First, we consider that the source (a directional beam pointed into the ice) is replaced with an effective, isotropic source located  below the surface at \(z=-l_{\mathrm{eff}}\). Then, the boundary condition with \(h=2l_\mathrm{eff}(1+R)/3(1-R)\) reads

\begin{equation}
    \phi(r,t) = \frac{2l_{\mathrm{eff}}}{3}  \frac{1+R}{1-R}\frac{\partial \phi}{\partial z}\ ,
    \label{eq:bcond}
\end{equation}

\noindent where \(R\) is the average Fresnel reflection as defined in \cite{haskell1994} and take the value of 0.3548 for diffuse light emerging from ice with a refractive index of 1.31. Employing the method of images and integrating eq. \ref{eq:bcond} in an analog fashion to \cite{bryan1890}, the mirror source distribution emerges with two new terms: Another effective source at \(z=l_{\mathrm{eff}}\) above the surface, and a damped line of sinks that partially counteracts the mirror source:

\begin{align}
   \phi (\rho,t,z=0) =   &\ \phi_+(l_{\mathrm{eff}}) + \phi_+(-l_{\mathrm{eff}})
     \nonumber\\
     & - \frac{2}{h}\int_0^\infty e^{-l/h}\phi_{+}(-l-l_{\mathrm{eff}})dl ,
    \label{eq:theo:bounded}
\end{align}

\noindent with the coordinates \(\rho,z\) as defined in Figure \ref{fig:setup}. Note that for total internal reflection of all radiation, i.e. \(R=1\), the sink term disappears as one would expect. A detailed derivation of this equation and its implication on retrieving scattering and absorption coefficient as well as volume depth can be found in \cite{allgaier2021}. For an appropriate measurement, where a short pulse of light is injected into the medium at \(\rho=0,z=0,t=0\) and a detector is placed far away at \(\rho\), this formula can be fitted to the measured fluence rate \(\phi (\rho,t)\), with scattering coefficients \(\sigma_{\mathrm{eff}}\) and absorption coefficient \(\sigma_{\mathrm{abs}}\) (or the respective length scales \(l_{\mathrm{eff}},l_{\mathrm{abs}}\)) as free parameters.

While this manuscript's focus is on measuring optical properties of bubbly glacier ice, the principle of the technique applies to any scattering medium. Here, we measure only effective scattering coefficients, which apply to an isotropic scattering process. The advantage of this approach is that assumptions about the nature of the scattering process and its distribution of scattering angles is not required. However, a discussion of the relationship between effective scattering coefficient and the physical mean-free path can give some context for different scattering media. For cold media - glacier ice, sea ice and snow - there are important differences for how the effective scattering length relates to established microscopic quantities. In bubbly ice, previous work assumed that the ice is solid with the refractive index being purely the one of ice. Scattering being dominated by bubbles of some average size and distance, which govern the mean-free-path of the physical scattering process \citep{price1997}. Bubbles (similar to grains in snow) are assumed to be large compared to the wavelength of light. In reality, this scattering process is far from isotropic. \citep{askebjer1997,price1997,ackermann2006} report a value of the average cosine of the polar scattering angle \(g=|\cos \theta|\approx 0.75\) for the case of Mie scattering on spherical bubbles, \cite{kokhanovsky2004a} report a value of 0.78 for the geometric contribution of scattering by ice spheres. It has been shown that any scattering problem in radiative transfer can be described through an effective, isotropic scattering process \citep{kokhanovsky2004a}, an approach that has been adopted by \cite{askebjer1997} and \cite{ackermann2006} as well. \(g\) then provides a convenient relationship between the physical mean free path \(l_{\mathrm{mfp}}\) and the effective scattering length \(l_{\mathrm{eff}}\), which describes an effective, isotropic process:

\begin{equation}
    l_{\mathrm{eff}} = \frac{l_{\mathrm{mfp}}}{1-g}.
\end{equation}

The scattering length is the inverse scattering coefficient, \(l_{\mathrm{eff}}=1/\sigma_{\mathrm{eff}}\) and therefore

\begin{equation}
    \sigma_{\mathrm{eff}} = (1-g) \sigma_{\mathrm{sca}} .
\end{equation}

While scattering in near-surface ice is dominated by bubbles instead of cracks, dust and crystal boundaries \citep{price1997,dadic2013}, these contributions may be significant close to the surface. For sea ice, liquid brine or salt crystals contained in the ice complicated things further since they can display a wide range of values for both their scattering cross section as and \(g\) \citep{light2004}. The advantage of measuring the effective scattering coefficient directly is that no assumptions about \(g\) or the nature of the underlying scattering processes need to be made.

Although the refractive index of glacier ice (in contrast to water ice containing no bubbles) depends on the volume fraction occupied by bubbles and therefore density, the same is true for the absorption coefficient \citep{dombrovsky2020}. With the absorptive component of the diffusion solution in Eq. \ref{eq:diffsolution} defined as \(\beta=c_0\sigma_{\mathrm{abs}}/n\), the density dependence of absorption coefficient and refractive index cancel.

%For the purpose of detector calibration at AMANDA and IceCube, the diffusion model was of limited use as deep ice at the South Pole does not strongly scatter light, putting the detectors effectively too close into the scattering-near-field.

%For snow, the situation is more complicated. Snow consists of air ice crystals of complicated shape. The shape, size and density of snow crystals influences the effective refractive index of the medium, scattering and absorption coefficient. For an in-depth treatment of these phenomena, see \citep{warren1980a,kokhanovsky2004,libois2013}. Given that different physical conditions possible in snow can yield identical effective absorption and scattering coefficients, we believe that measuring these effective properties in the first place offers a more palatable approach to describing radiative transfer in snow.

\section{Experimental Methods}

\begin{figure}
    \centering
    \includegraphics[width=0.45\textwidth]{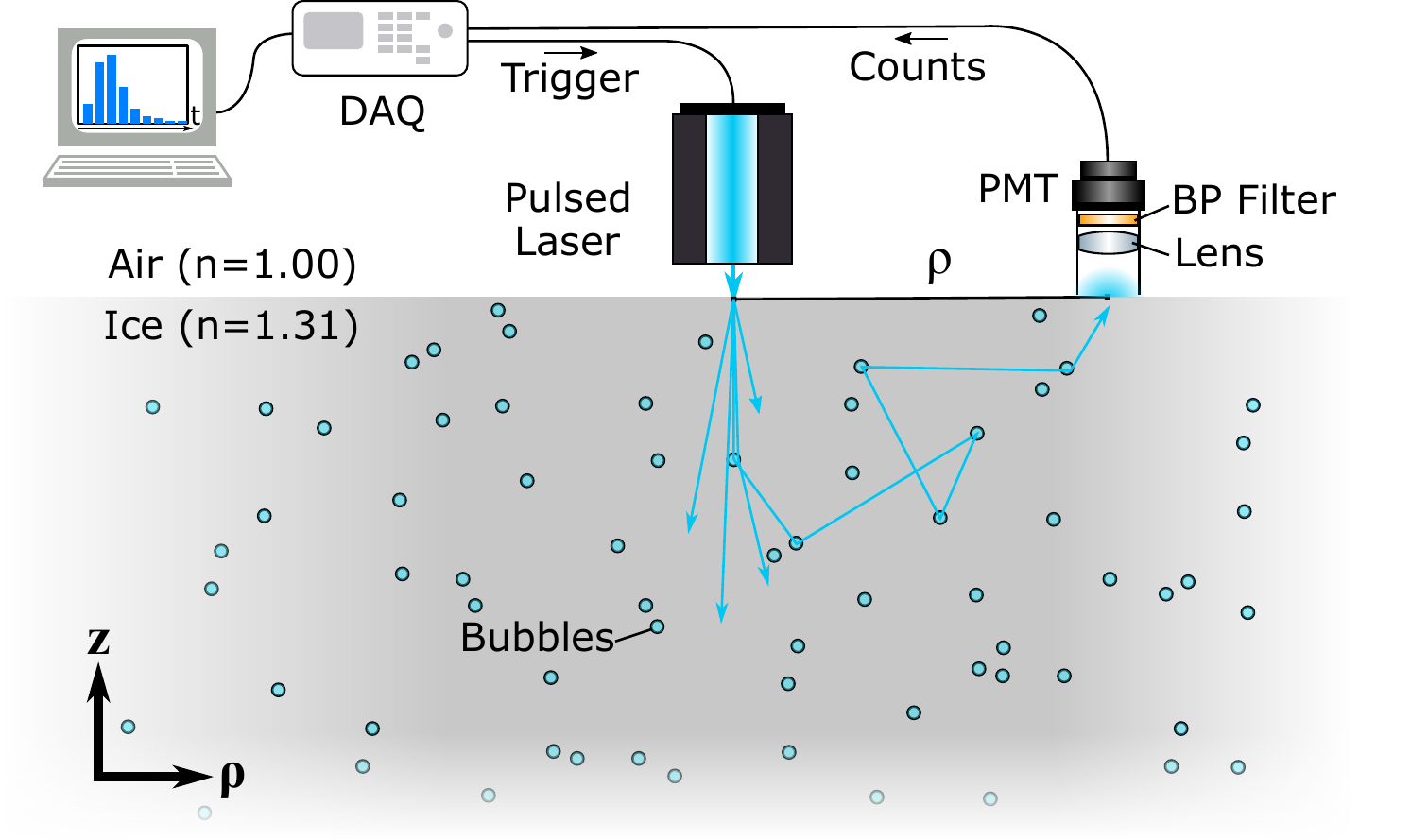}
    \caption{The experimental setup is at core similar to a photon counting LiDAR. The light source is a diode laser module, which emits pulses when triggered by the data acquisition module (DAQ). A delayed trigger pulse is used to gate the DAQ's counting circuit, which therefore only counts the events received from the photon counting detector in a discrete time interval after pulse emission. The arrival-time histogram is integrated in the LabView computer interface by shifting the delay between laser trigger and counter gate in constant time intervals. The backscattered light is collected onto the detector using a 50mm-focal length lens and cleaned of most background light with a 10nm-bandwidth bandpass filter. Both optical elements are inside a lens tube, which provides shielding from background. The photon counting detector is a photo-multiplier tube (PMT) with an active area 5mm in diameter.}
    \label{fig:setup}
\end{figure}

To measure the absorption and effective scattering coefficient of glacier ice, it is necessary to measure the fluence rate and fit the result with the diffusion theory from Eq. \ref{eq:theo:bounded}. The fluence rate describes how much energy flows through a unit area at arbitrary orientation per unit time. In our setup as shown in Figure \ref{fig:setup} we inject a laser pulse vertically into the ice. The detector is placed in the scattering-far-field, in this specific case about 1.8m away from the laser, which is of the order of \(\approx 30l_{\mathrm{eff}}\). A realistic detector has a finite area defined by the area imaged by the collection optics. Therefore, the appropriate function to use for a fit is the fluence rate integrated over this area, which possesses units of energy per unit time. Realistically, placing such a detector in the scattering-far-field introduces many orders of magnitude of loss \citep{allgaier2021}, which can only be overcome by using photon-counting detectors, a feature absent from most commercial LiDAR systems. At the same time, these detectors provide excellent temporal resolution.

While pulsed laser sources, photon counting detectors and the necessary timing electronics are all commercially available, not all varieties are affordable or appropriate. As shown in \cite{allgaier2021}, the expected duration of the fluence rate distribution is between 100ns and 1000ns, which implies that a 10ns-duration laser pulse is significantly shorter than the overall distribution, rendering it effectively "instantaneous". The same is true for the employed electronics: The measurement can be implemented using the counter circuit of many data acquisition systems with ns-resolution. This option can offer sufficient resolution at comparatively lower cost.

We employ three different nanosecond-pulsed diode lasers (\textit{Thorlabs NPL41B, NPL52B and NPL64B}) set to a pulse duration of 18ns and a repetition rate of 1MHz. All measurements are repeated with three different wavelengths. 405nm is close to the absorption-minimum of ice \citep{warren2019}. 520nm is close to the laser used by ATLAS onboard ICESat-2 (532nm). The third laser at 640nm serves as a reference. In transmission measurements, it has been assumed that the absorption coefficient of ice is mostly independent of LAPs at wavelengths longer than 600nm. With the laser at 640nm we can test this assumption. At a pulse duration of 18ns the laser emits reasonable pulse energies (70pJ at 405nm, 57pJ at 520nm and 95pJ at 640nm) without drastically diminishing the temporal resolution dominated by other components.

The detector is a photo-multiplier tube (PMT) module (\textit{Hamamatsu H7421-50}) with an active area 5mm in diameter. 
Scattered light is collected by a lens with NA=2 (25 mm diameter, focal length 50 mm) and spectrally filtered with a 10nm-bandwidth bandpass filter centered on each laser wavelength (\textit{Thorlabs}), both placed in lens tubes which do not require alignment in the field. Compared with a lens as well as suing a lens tube provides additional shielding from background light. The detector has a quantum efficiency of 12\% at 640nm that outputs TTL pulses with a rise time of <1.5ns upon detecting a photon. The lens is set to focus collimated light onto the active area, effectively imaging an area 25mm in diameter. Even with the bandpass filter in place, the photon-counting PMT module is saturated and possibly destroyed in daylight conditions. Therefore, measurements must be conducted after sunset.

In our setup, the time-resolved measurement is implemented using a data acquisition module (DAQ - \textit{National Instruments USB-6361}). We use one counter circuit to generate a pulse train with a repetition rate of 1MHz to trigger the laser. A second counter circuit is used to generate a pulse train that is synchronized to the first one except for a time delay. This second pulse train acts as a time gate for the third circuit which is used as an even counter. When gated, the counter circuit only counts events that arrive in the time window defined by the duration of the gate pulse. In this configuration, the fluence rate \(\phi(\rho,T)\) can be measured as a function of delay \(T\) from the laser pulse launch time in a time interval of 20ns, which is approximately equal to the input laser pulse duration. The measured time-of-flight (ToF) histogram is the average fluence rate in the measured time interval of 20ns length. The complete ToF histogram, is measured by stepping the delay \(T\) between the laser trigger and the counter gate pulse trains. With the temporal resolution of 20ns (limited by the gate duration) and a pulse repetition rate of 1MHz, we can resolve ToF distributions as short as a few time bins (one time bin being the gate duration of 20ns), and as long as the time between two pulses (1000ns). Since the laser is triggered externally using the DAQ, these parameters can all be easily adjusted in the field. With the laser repetition rate of 1MHz, the time between consecutive pulses is 1000ns, much larger than the dead time of the detector. In addition, the strong loss between the laser and detector guarantees that the number of detected photons per pulse is \(\ll1\). This guarantees that the detector response in linear.

The histograms are evaluated with the native bin size of 20 ns as defined by the gate pulse duration of the DAQ. The last 5 bins (t=900...1000 ns), which we can assume to only contain background, are averaged and the resulting background contribution is subtracted from all bins. Given laser-detector separation \(\rho\) (measured with a tape measure), Eq. \ref{eq:theo:bounded} is then integrated over the time bin interval and fitted to the measured ToF histogram (Fig. \ref{fig:crook} solid lines).

\begin{figure}
    \centering
    \includegraphics[width=0.45\textwidth]{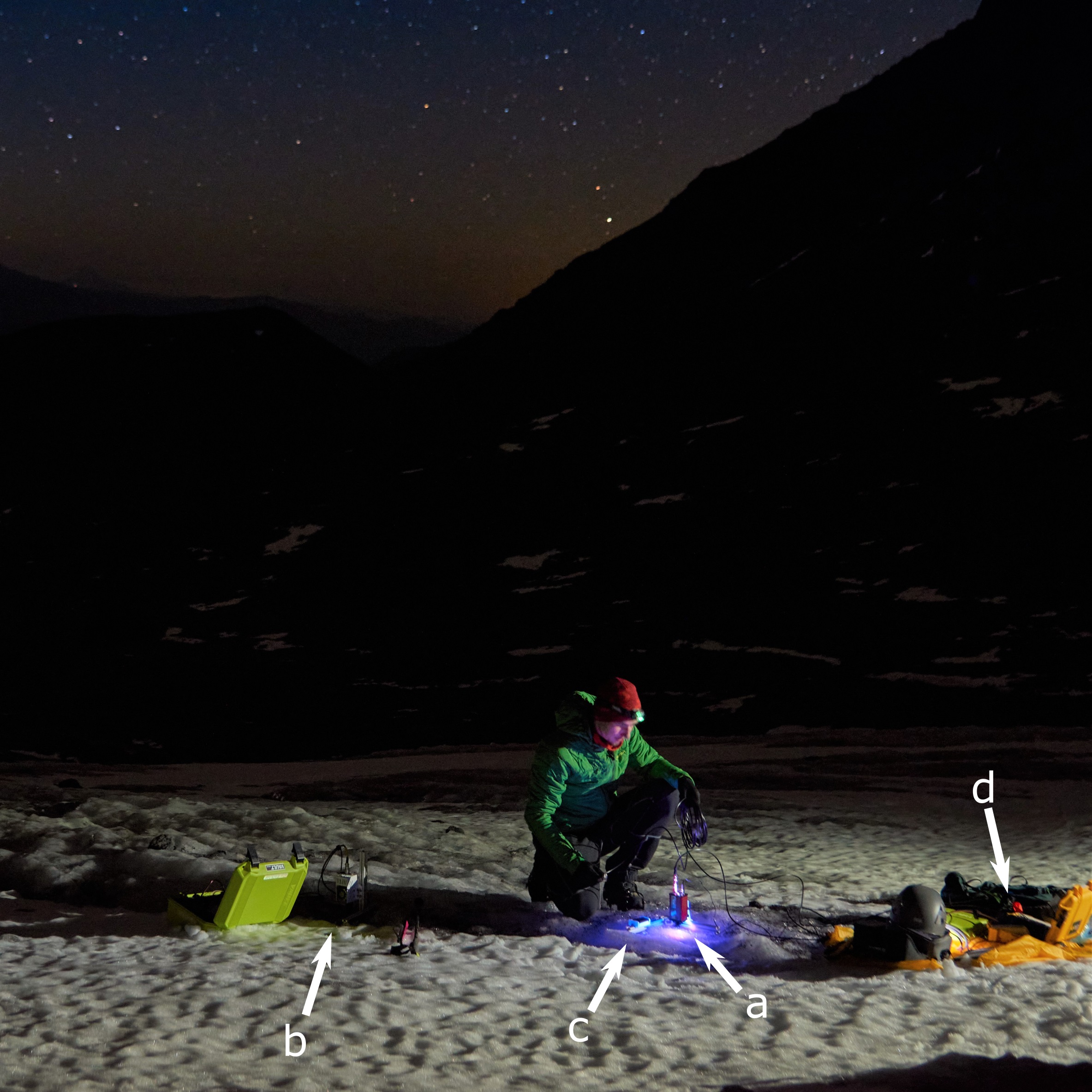}
    \caption{Backscattered blue light can be seen right below the laser (a). The PMT is positioned to the left (b).  A GPS receiver is co-located with the laser (c). The DAQ remains in its hard case (d).}
    \label{fig:setup2}
\end{figure}

%The overall temporal resolution of the setup in this configuration is \(\Delta t = \sqrt{(20ns)^2 + (18ns)^2} = 27ns\).
The setup does not need any alignment in the field and can be powered using a 200Wh 12V battery for the DAQ and laser and a 50Wh battery for the PMT (\textit{Goal Zero}), which provide a runtime of roughly 10 hours on the ice, rechargeable using portable solar panels. The DAQ is controlled using a LabView interface run on a tablet computer. The combination of being battery powered, using low-power (effectively class-1) lasers and transporting the instrument on foot ensures minimal environmental impact and enabled us to work in designated wilderness areas. Figure \ref{fig:setup2} shows the instrument in operation on Collier Glacier in the Three Sisters Wilderness, Oregon.

In general, there are internal electronic delays between the different counter circuits, between laser trigger and optical output, as well as physical delays due to the length of cables. All of these need to be measured and subtracted from the measured ToF distribution to extract the "actual" time of flight. This is done by placing laser and detector close together and less than 30cm away from a solid surface (scattering card or a wall) and measuring a ToF histogram. The actual time of flight would be the physical path from the laser to the surface plus the distance from the surface to the detector, in this case <2ns, unresolvable by the system. The peak at the measured distribution indicates the internal delay, which can then be subtracted from all measured ToF distributions. With the cables used here, the internal delay is 130ns for all measurements taken on Collier Glacier and 190ns for the calibration measurements taken on Crook Glacier.

%While using a more powerful laser with higher pulse energy or a time tagger would reduce measurement duration and improve temporal resolution, they would also be much more expensive, with the current implementation costing around \(\$10,000\).
A complete list of parts required to built this setup (a cost of roughly \$7,000) and further explanation of the timing circuitry can be found in the appendix.

\section{Study area}
\label{sec:area}

\begin{figure}
    \centering
    \includegraphics[width=0.45\textwidth]{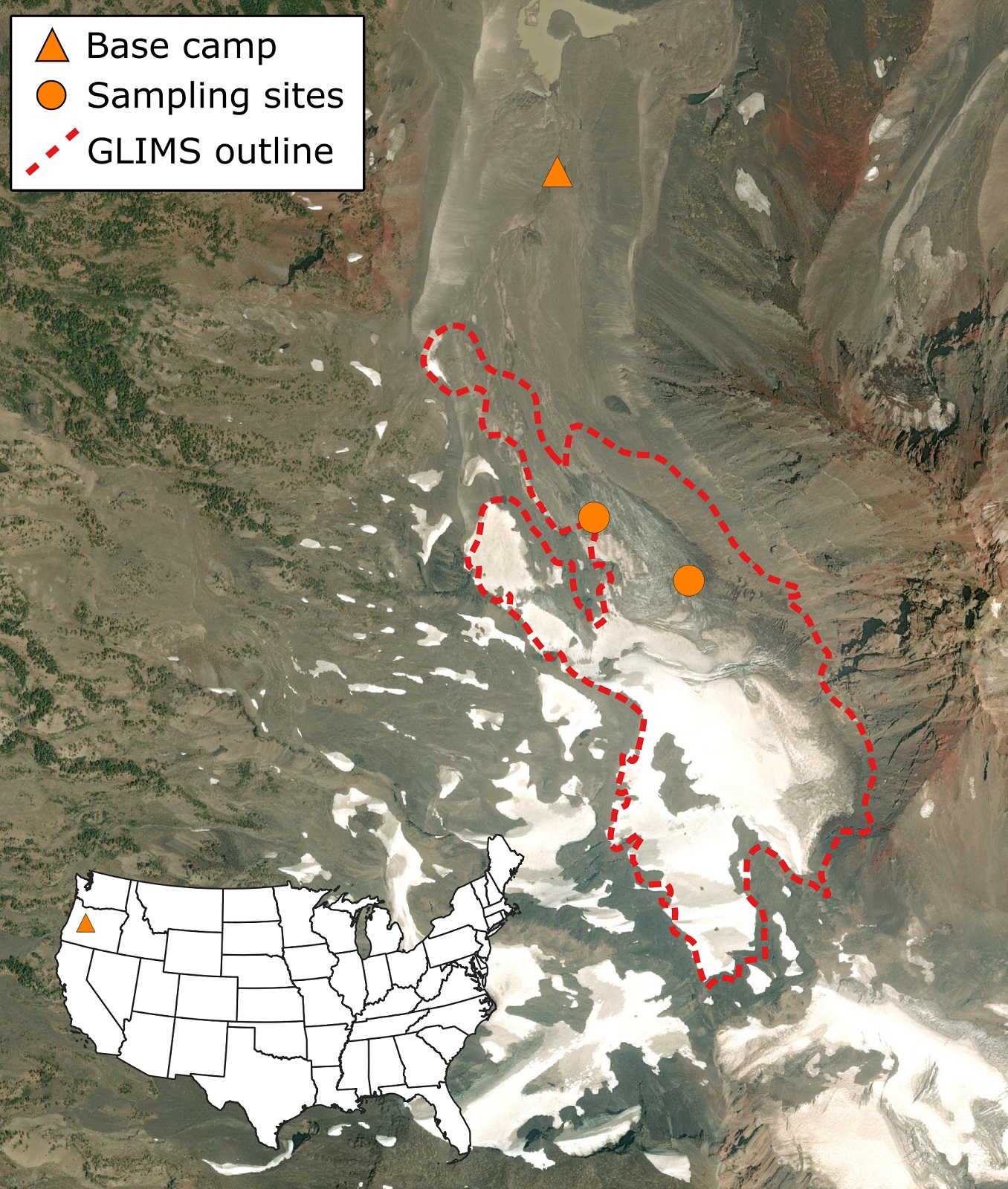}
    \caption{The study area on Collier Glacier, Three Sisters Wilderness, Oregon. (a) Two sites were sampled: One close to the northern terminus, and one in the middle of the glacier. (b) Site 2, close to the middle of the glacier below the ice fall which roughly marks the average ELA.}
    \label{fig:site}
\end{figure}

Field work was conducted on Collier Glacier located in the Three Sisters Wilderness, Oregon, USA (see Figure \ref{fig:site}). Collier Glacier, is a temperate valley glacier in Oregon with an elevation of between 2300 m and 2700 m and an area of roughly 0.7km\(^2\) \citep{mountain1978,beedlow2011}. Data was collected at two sites (44.169444\(^{\circ}\)N, 121.788256\(^{\circ}\)W, 2324 m and 44.167509\(^{\circ}\)N, 121.784424\(^{\circ}\)W, 2406 m, respectively) on September 24\textsuperscript{th} and 25\textsuperscript{th} during which the glacier was mostly free of snow, with only the crevasses and upper-glacier being covered by a recent layer of snow (the measurements were made on bare ice).

Data collection was performed during two nights starting about 1 hour after sunset (20:00) when the background light level had subsided sufficiently as to not saturate the PMT. At each site, 8-10 measurements at each wavelength (405, 520 and 640 nm, 30 measurements total at each site) were taken over the course of 3-4 hours each night, with the initial setup taking not more than 15 minutes. For these measurements, the laser and detector were moved to different positions on an area of 4\(\times\)6m, where the distance between laser and detector was kept between 1.5 m and 1.9 m, determined with a tape measure. All measurements were completed for one wavelength before moving on to the next. This order is preferable over taking three measurements at different wavelengths at the same location, because the laser head needs several minutes to warm up and achieve constant output power.

In addition to the measurements on Collier Glacier, a series of measurements aimed at determining optimal distance between the laser source and the detector was conducted on September 16\textsuperscript{th} on the nearby Crook Glacier (44.079917N, 121.698415W, 2506 m), which unlike Collier Glacier widely lacks reference data from previous research but offers far easier access. On Crook Glacier, the laser mount was kept in the same position throughout the night, swapping the laser heads in place to ensure that measurements at different wavelengths were comparable. The detector was moved to distances from the laser between 0.8 m and 2.5 m with the goal of establishing a useful distance range in which ToF histograms could be measured with a reasonably well-resolved peak (histogram occupying at least 5-6 bins) and sufficient signal-to-noise ratio (SNR - at least 1000 counts per bin after background-subtraction).

\section{Results}

\subsection{Data Analysis}
To analyse the measure ToF histograms, Eq. {eq:theo:bounded} is used to perform a least-square fit. The theoretical model is amended by a multiplicative scaling constant \(C\) as well as a timing offset:

\begin{equation}
    \phi_{\mathrm{fit}} = C \cdot \phi (\rho,t+ \Delta t,\sigma_{\mathrm{eff}},\sigma_{\mathrm{abs}}) 
    \label{eq:fit}
\end{equation}

\noindent with \(C, \sigma_{\mathrm{eff}},\sigma_{\mathrm{abs}},\Delta t\) as fit parameters. While in theory, the absolute arrival time offset \(\Delta t\) should be zero, we can only characterize it with the limited resolution of the time-resolved counter (20ns). We hence include it as a fit parameter. Eq. \ref{eq:fit} is additionally integrated over each 20-ns bin in the measured ToF histogram (See Fig. \ref{fig:crook}). This integrated value is what is compared to the data in the least-square fit routine.

\subsection{Characterization at optimal laser-detector separation}

Measurements taken on Crook Glacier indicated that the appropriate distance between laser source and detector was between 1.5 m and 2 m. Below 1.5 m, the peak of the ToF histogram was poorly resolved due to insufficient temporal bin width, especially for the red 640 nm-laser, where absorption is stronger. Above 2 m, ToF histograms were increasingly hard to measure, with the required integration time exceeding 10 minutes accompanied by poor SNR. Note that the laser mount was left in place to keep alignment constant, which is necessary to compare measurements at different wavelengths.

\begin{figure}
    \centering
    \includegraphics[width=0.45\textwidth]{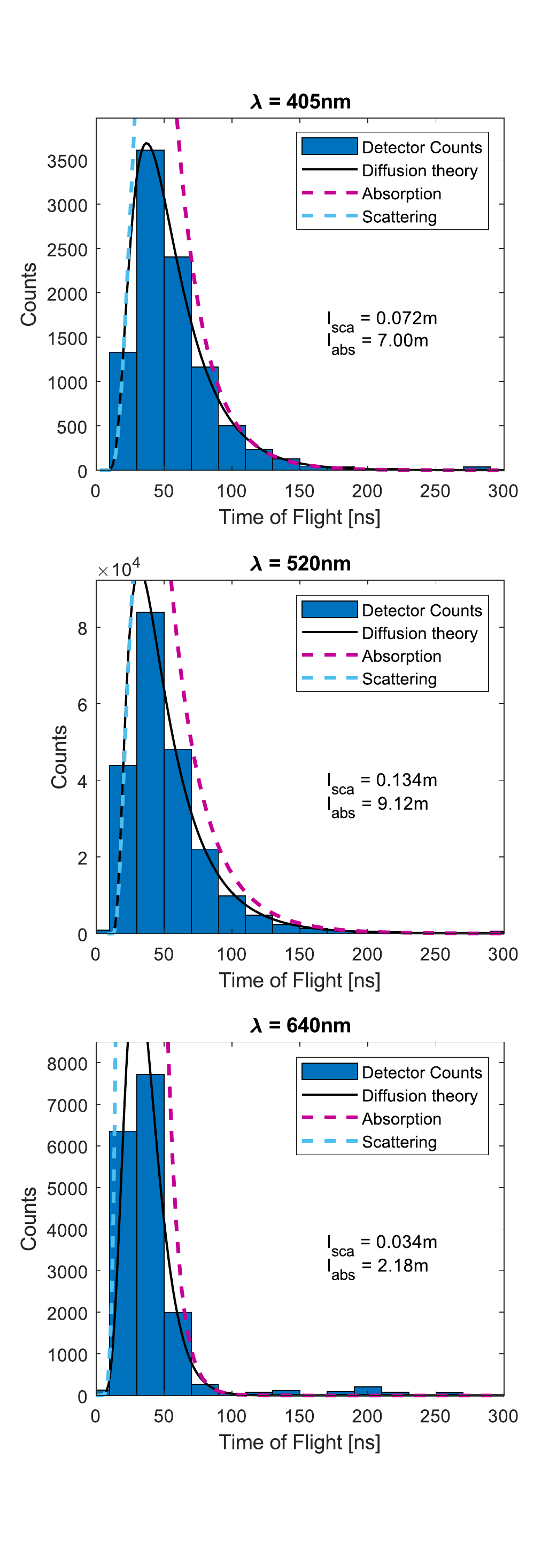}
    \caption{Arrival-time histograms for the three different wavelength as recorded on Crook Glacier. Blue bars represent the background-subtracted data. The solid black line is the fitted diffusion curve, from which absorption and scattering coefficient are extracted. The cyan and magenta dashed lines are the diffusion curve with the absorptive and scattering component, respectively, turned off.}
    \label{fig:crook}
\end{figure}

For the ideal separation, we show three ToF histograms recorded at a distance of 1.4 m in Figure \ref{fig:crook}.
The figure also shows that same result of the fit with the absorptive and scattering component turned off, respectively (cyan and magenta dashed lines, respectively), re-scaled for better visibility. This shows that the rising edge of the distribution is governed mostly by the scattering coefficient, and the falling edge and slow fall-off is where the absorption coefficient is extracted. With identical laser position, all three measurements yield scattering coefficients of 14(\(\pm\)4)\(m^{-1}\), 7(\(\pm\)2)\(m^{-1}\) and 29(\(\pm\)7)\(m^{-1}\), respectively for 405nm, 520nm and 640nm, where we assume a relative uncertainty of 25\%, derived from the reliability of the curve fit on simulated data \citep{allgaier2021}. The absorption coefficients extracted from the fitted curve (0.14(\(\pm\)0.04)\(m^{-1}\), 0.11(\(\pm\)0.03)\(m^{-1}\) and 0.5(\(\pm\)0.1)\(m^{-1}\), respectively for 405 nm, 520 nm and 640 nm) are in the range of literature values \citep{warren2008}.

It can, however, be seen that the peak is poorly resolved at 640nm, which can be attributed to the strong absorption at that wavelengths, resulting in a truncated peak structure, which doe not change as a function of distance. We would require higher temporal resolution to resolve the peak at 640 nm. At smaller separation, the peak onset is poorly resolved at all wavelengths. At separations larger than 2 m, histograms are exceedingly hard to measure, requiring large integration times. Therefore, we conclude that the ideal distance between detector and laser source for the study area is between 1.5, and 2 m. Overall, our instrument successfully retrieved both scattering and absorption coefficients through successful curve fits for 405nm and 520nm, where both the peak onset and the tail of the distributions are well resolved. Shortcomings particularly in the parameter retrieval at 640nm will be discussed in the following section.

\subsection{Statistical independence of scattering and absorption coefficients}

\begin{figure}
    \centering
    \includegraphics[width=0.5\textwidth]{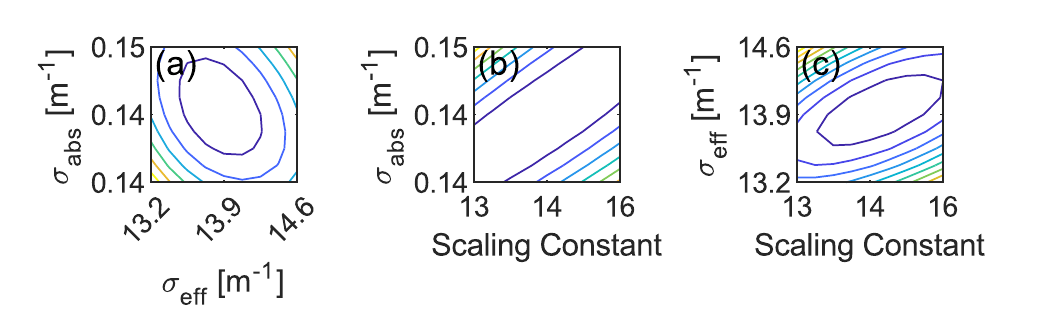}
    \includegraphics[width=0.5\textwidth]{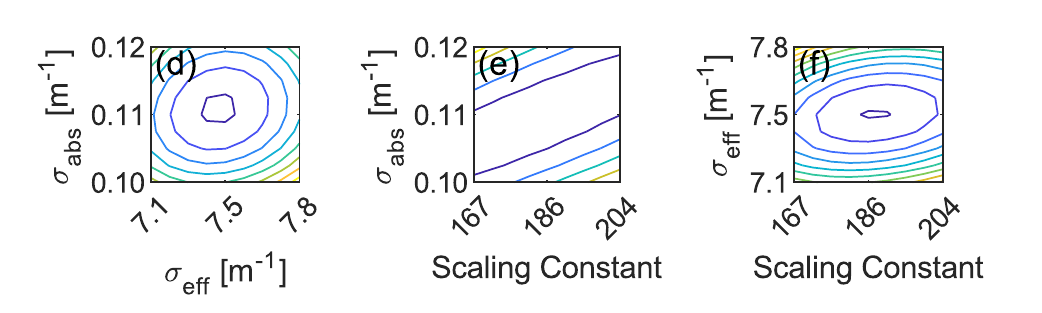}
    \includegraphics[width=0.5\textwidth]{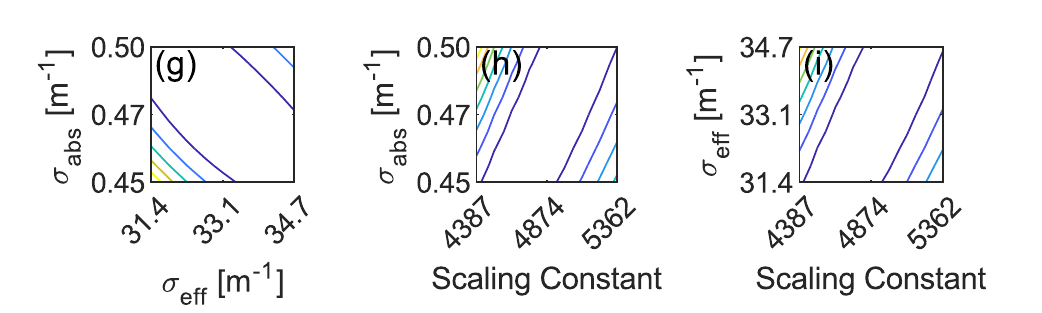}
    \caption{Sum of square errors (arb.u.) between fit and data for the parameter space around the best-fit-values extracted from the histograms shown in Fig. \ref{fig:crook}. Panels (a,d,g) show the sum of least squares for combinations of different values for the effective scattering and absorption coefficients with the other two parameters kept constant for the histograms recorded at 405, 520 and 640nm, respectively. Panels (b,e,h) show the same data for the scaling constant and absorption coefficient. Panels (c,f,i) are for the combination of scaling coefficient and scattering coefficient.}
    \label{fig:chi2}
\end{figure}

To judge the quality of the obtained curve fits, we vary each of the four fit parameters for the data shown in Fig. \ref{fig:crook} around the respective values obtained from the fit and calculate the \(\chi^2\) (sum of the square deviations). The resulting \(\chi^2\)-landscapes are shown in Fig. \ref{fig:chi2}. The panels in the first column (a,d,g) show the \(\chi^2\)-landscape for the combination of scattering and absorption coefficient (with the other two fit parameters held constant) for the three previously shown fits. For the data obtained at 405nm and 520nm (panels a and d), the \(\chi^2\)-landscape constrains the fit values in both directions. The 405-nm data exhibits a slight anti-correlation. The data for 640nm on the other hand (panel g) shows this anti-correlation in a much more severe way, as the absorption and scattering coefficients are not independently constrained within the plotted window of \(\pm5\%\) around the fit values. This implies, that we cannot retrieve independent value for the absorption and scattering coefficients from such poorly resolved data. The second column (panels b, e, h) show the same plot but for the combination of absorption coefficient and the multiplicative scaling factor. These tend to show anti-correlation, however, the same is true for the scaling factor and scattering coefficient, which are shown in panels (c, f, i). Therefore, the scaling factor correlates identically with both the absorption and scattering coefficient, implying that all are well constrained in a fit where they are all free parameters.

Particularly the fit behavior for the two important optical parameters in question here, scattering and absorption coefficient, is well constrained and therefore the fit should yield an independent value for both. Therefore, weak correlations like what the one shown in Fig. \ref{fig:chi2}a will not manifest. This is not true for the data taken at 640nm, where the fit has to be assumed to be less reliable. This will become evident as we examine correlation between many different data points in the following.

\begin{figure}
    \centering
    \includegraphics[width=0.5\textwidth]{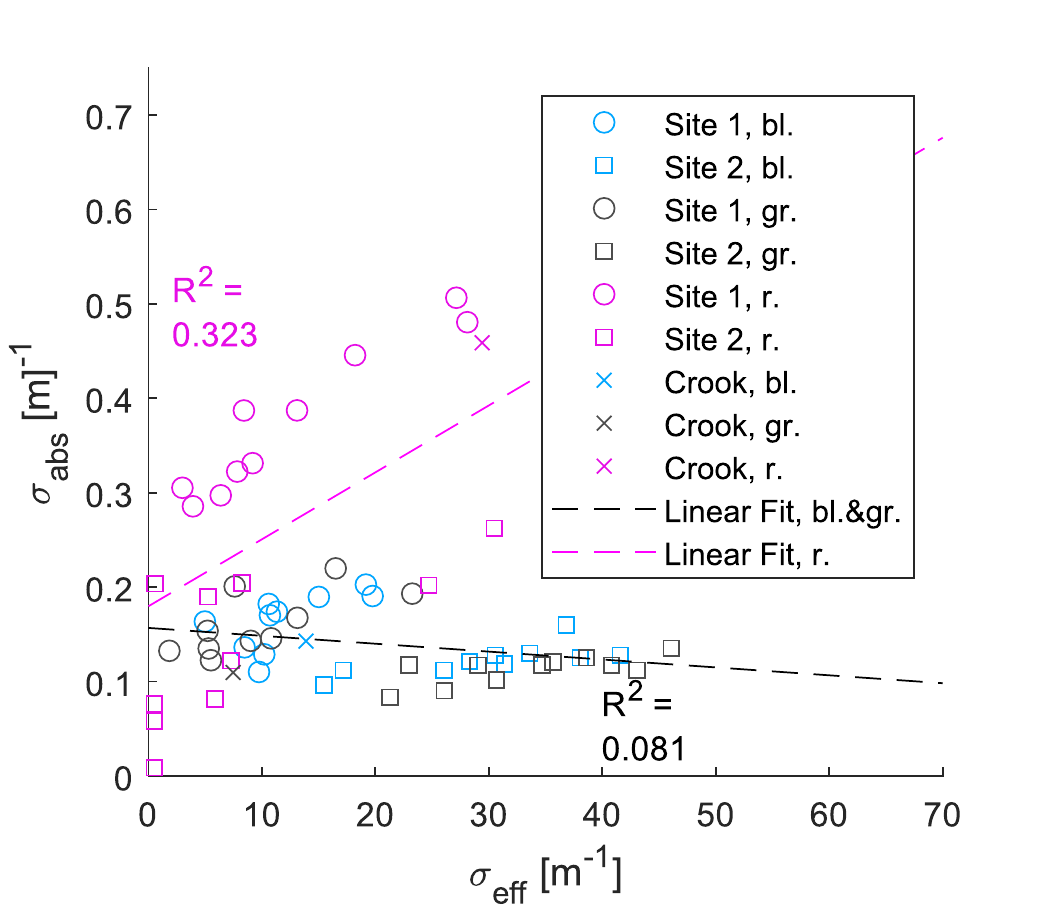}
    \caption{All combined data points for scattering and absorption coefficients from both Crook and Collier glacier at all three wavelengths. The dashed black line represents a linear fit to blue and green data ($R^2=0.081$), which illustrates the statistical independence of the retrieved parameters. The magenta dashed line ($R^2=0.323$) represents the linear fit to red data.}
    \label{fig:combined}
\end{figure}

In practice, there should be no correlations between the physical scattering and absorption coefficients found in typical surface glacier ice, where scattering is mostly due to bubbles, and absorption is dominated by LAPs. To test this assumption, a large and diverse data set is needed to ensure that there are no such correlations in the underlying coefficients. For this purpose, we combine the data measured at both sites on Collier Glacier as well as the site on Crook Glacier. Given the \(\chi^2\)-study, we interpret the data measured at 405nm (blue) and 520nm (green) separately from the one taken at 640nm (red). The fits for all histograms were unsupervised using identical initial parameters.  A scatter plot of the successful measurements is shown in Figure \ref{fig:combined}.

To establish whether the extracted value pairs are statistically independent or not, we perform a linear fit and calculate the R\(^2\) of the distribution. For blue and green data combined, we find R\(^2\)=0.081, while it is 0.323 for the red data. This demonstrates that a linear best fit represents the distribution no better than a simple average in the blue and green, however, there is a clear correlation between absorption and scattering in the red data set. We therefore conclude that this method can indeed provide an independent measurement of scattering and absorption coefficient as long as ToF histograms are well resolved, as is the case at 405 and 520nm. However, this is not true for data taken at 640nm, where insufficient resolution manifests both in a larger spread of the measured values as well as clear correlations between the two coefficients.

\subsection{Results from Collier Glacier}

\begin{figure}
    \centering
    \includegraphics[width=0.45\textwidth]{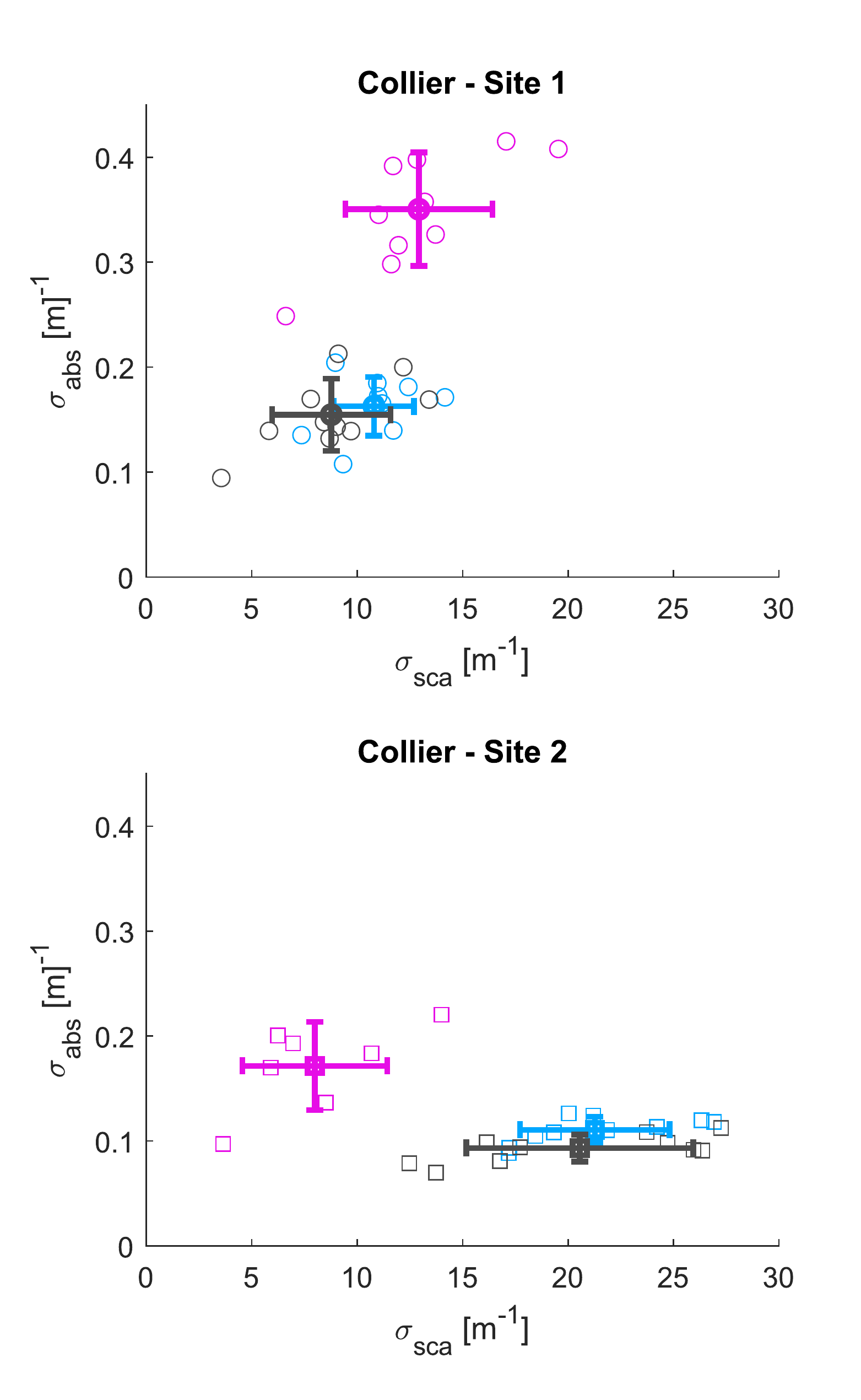}
    \caption{Scattering and absorption coefficients from the two sites on Collier Glacier with black, cyan and magenta representing data taken at 405nm, 520nm and 640nm, respectively. Thick symbols are the mean scattering and absorption coefficient for each wavelength, and the error bars representing the standard deviation of each distribution.}
    \label{fig:collier}
\end{figure}

The extracted scattering and absorption coefficients for the two sites are shown in Figure \ref{fig:collier}. Thicker symbols represent the mean coefficient pair at each wavelength, with the error bars representing the standard deviation. We use the standard deviation as a conservative estimate of the measurement uncertainty, as the observed spread encompasses both uncertainty from the model fit and natural variations of the glacier ice properties. First, it can be seen that the scattering coefficients for the green and blue wavelength, for which the peaks of the ToF distribution are generally well resolved, are comparable at each of the two sites. The red scattering coefficients seems at site 2 to be significantly different from the others. This is also true when we average the coefficients over both sites (20.9(\(\pm\)1.5)m\(^{-1}\), 22.2(\(\pm\)3.0)m\(^{-1}\) and 11.9(\(\pm\)2.6)m\(^{-1}\) at 405 nm, 520 nm and 640 nm, respectively), where the stated uncertainties are standard errors. While the values at 640nm differs significantly from the others, it is still within a 2\(\sigma\) uncertainty range. That said, we emphasize that the scattering coefficient measured at 640 nm suffers from poor temporal resolution, which implies that the scattering coefficient may be underestimated (compare site 2).

The extracted absorption coefficients provide a more uniform picture. Average values at 405 nm and 520 nm are almost identical and, as expected, red absorption is stronger than green and blue. The absorption coefficients (10\(^{-1}\mathrm{m}^{-1}\)) are at the upper end of the range of previously published values \citep{warren2008} (c.f. Figure 12 of \cite{cooper2020}), suggesting that the ice we measured at Collier Glacier contained substantial concentrations of BC or other LAPs. Such large coefficients and the fact that the absorption is almost identical for the green and blue part of the spectrum agree with observations made on icebergs and glacier ice contaminated with algae and organic LAPs \citep{warren2019b}. The fact that the absorption coefficient at Site 2 is lower than it is on Site 1 would also indicate a larger albedo, consistent with previous observations on Collier Glacier, where the broadband albedo was found to be lower at the terminus \citep{mountain1978}.

While previous work often considered the absorption coefficient at wavelengths >600 nm to be independent of LAPs, it is apparent from simulations of snow albedo that this is only true for small concentrations \citep{warren1980b}. However, due to its smaller scattering coefficient, ice albedo would depend more strongly on the absorption coefficient and LAPs. We will elaborate on this aspect in the next section for the example of BC.

To summarize, the observed absorption coefficients are consistent with glacier ice contaminated by LAPs, which is to be expected for any glacier outside of the Arctic or Antarctic.

\subsection{Outlook: Penetration depth, Albedo and LAP concentration}

Our measured scattering and absorption coefficients are relevant to a number of different applications, including inference of laser altimeter penetration depth, albedo, and LAP concentrations. Firstly, the effective scattering length represents the depth at which the direction of scattered photons has been fully randomized. In the limit of small absorption, this is the approximate depth at which a maximum of backscattering to the surface occurs. In the context of laser altimetry or LiDAR mapping, this can be interpreted as the penetration depth. Averaged over both sites, the scattering length at 520 nm (close to common wavelengths for LiDAR systems) is 6.9 cm, meaning that LiDAR measurements taken at nadir need to account for such a bias. Our measurements can be considered a first-order approximation of this bias, but more work is needed to quantify the magnitude and variability of this bias over different surfaces and to understand effects such as pulse broadening.

%Secondly, it has been shown that the diffuse (spherical) albedo is fully described by effective bulk scattering and absorption coefficients like the ones measured here \citep{kokhanovsky2004}. In the limit of \(\sigma_{\mathrm{eff}}\gg\sigma_{\mathrm{abs}}\), the albedo \(\alpha\) then reads

%\begin{equation}
%    \alpha = 1 - 4\sqrt{ \frac{\sigma_{\mathrm{abs}}}{3 \sigma_{\mathrm{eff}}} }.
%\end{equation}

Secondly, it has been shown that the radiative transport for bubbly glacier ice can be framed in terms of effective bulk scattering and absorption coefficients \citep{dombrovsky2020} like the ones measured here . To this end, we first define the effective single-scattering albedo \(\omega = \sigma_{\mathrm{eff}} / (\sigma_{\mathrm{eff}}+\sigma_{\mathrm{abs}})\). Since the refractive index of ice \(n\) is larger than that of of air, total internal reflection occurs for upwelling radiation impinging on the air-ice-interface at angles larger than \(\mu_c=\sqrt{1-(1/n^2)}\). \cite{dombrovsky2020} then define the following parameters needed to express the reflectance function:

\begin{equation}
    \xi^2 = \frac{4}{(1+\mu_c)^2}\frac{1-\omega}{1-\omega \mu_c},\quad  \chi=\frac{\omega(1-R(1))}{1-\omega},\quad \gamma=\frac{1-R(1)}{1+R(1)},
\end{equation}

where \(R(1)\) is the Fresnel reflection coefficient at normal incidence. The plane albedo \(r_p\) then reads

\begin{equation}
    r_p(\mu_i) = R(\mu_i) + \frac{\gamma}{2n^2}\frac{\xi^2\chi}{(\xi+1/\mu_i)[\xi+2\gamma/(1+\mu_c)]}.
\end{equation}
\noindent where \(\mu_i\) is the cosine of the incidence angle.

\begin{figure}
    \centering
    \includegraphics[width=0.45\textwidth]{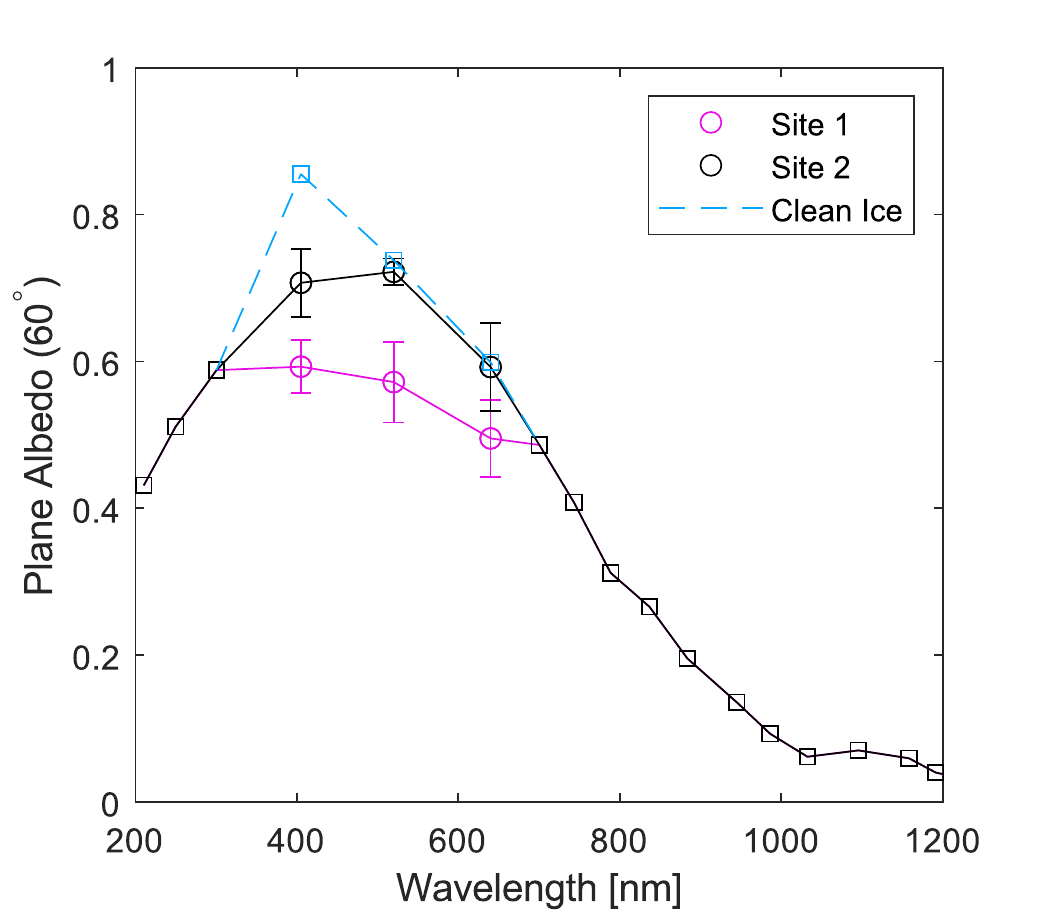}
    \caption{Spectral albedo on both sites as well as for clean ice. Data points at 405nm, 520nm and 640nm are calculated from measured scattering and absorption coefficients (open circles). Other data points are calculated from literature values for absorption and the averaged measured scattering coefficient (open squares). Solid lines represent linear interpolation as a guide to the eye. The dashed line was calculated using absorption coefficients for clean ice and the scattering coefficients from site 2.}
    \label{fig:albedo}
\end{figure}

Note that this equation takes into account a specular surface reflection. Using this expression we can calculate the plane albedo for any angle of incidence. We extrapolate the measured spectral components from 405-640 nm into the UV and IR and assume that the absorption coefficient is independent of LAPs outside of the visible range \citep{warren2008}. Then, we take the average of the measured scattering coefficients to calculate a complete spectral albedo curve. The results for an incident angle of 60\(^\circ\) are shown in Figure \ref{fig:albedo}, where reduction in spectral albedo throughout the visible spectrum from Site 2 to Site 1 is visible. We also show the spectral albedo assuming clean ice and measured scattering coefficients. To estimate a broadband albedo at each site, we calculate a weighted average of the spectral albedo, using a solar spectrum calculated with COART \citep{jin1994} as a weight. Values obtained in this way are 0.48\(\pm\)0.03 and 0.56\(\pm\)0.03 for Site 1 and 2, respectively, and 0.62 assuming clean ice.

Although established methods from reflectance spectroscopy provide measured spectral albedos, such techniques typically only answer the question \textit{what} the albedo is. In contrast, our method provides the underlying absorption and scattering length scales that explain \textit{why} albedo assumes a particular value, and decouples absorption and scattering properties, thus eliminating the need for further measurements of structure or composition.
%This is particularly important in the context of \textit{similarity}, where a particular value of albedo may be associated with different combinations of the scattering and absorption length scales (van de Hulst). Moreover, glacier ice albedo models are nascent relative to snow albedo models, and currently lack sophisticated physically-based treatments of ice physical properties that control glacier ice albedo. Such properties include the spatial and temporal distribution of mechanical fractures (cracks), air bubble size distributions, presence and thickness of granular near-surface ice (weathering crust), and presence of absorbing impurities such as black carbon, dust, micro-organisms, salts, and liquid water. Given that such information is not widely available at spatial scales over which albedo varies, one could extrapolate the measurements presented here to all ice of similar age, under the assumption that ice sharing a common history shares common first-level controls on its scattering and absorption length scales. The technique presented here could therefore efficiently improve on existing albedo models within a cost-effective measurement program.

Lastly, we use the measured absorption coefficient to estimate effective concentrations of LAPs. For this, we require a baseline absorptivity for clean ice. One such estimate by \cite{warren2006} gives a value of \(\sigma_{\mathrm{abs,clean}}\)=7.78\(\times\)10\(^{-4}\mathrm{m}^{-1}\), while \cite{picard2016} gives a more conservative estimate of \(\sigma_{\mathrm{abs,clean}}\)=1.9\(\times\)10\(^{-2}\mathrm{m}^{-1}\) at 405 nm. We will give results using both reference values. The density \(d\)=870kg/m\(^{3}\) has been measured using an ice screw with a diameter of 1.3 cm and length of 15 cm as a sampler. For this demonstration, we assume that BC is the only LAP, which implies that extracted concentrations are an upper bound. We assume an Angstrom coefficient of 1.1 and a mass absorption cross section of \(C_{BC}=\)8.9m\(^2\)g\(^{-1}\) at \(\lambda_{\mathrm{ref}}\)=450nm \citep{doherty2014}. Using these values the concentration can be calculated using

\begin{equation}
    C = \frac{\sigma_{\mathrm{abs}} - \sigma_{\mathrm{abs,clean}}}{dC_{BC} (\lambda/\lambda_{\mathrm{ref}})^{-1.1}}.
\end{equation}

For the two sites and using the clean ice absorption value from \cite{warren2006}, this yields BC concentrations of 18.9(\(\pm\)3.5) ppb and 14.1(\(\pm\)1.9) ppb, respectively. Using the clean ice absorption measured by \cite{picard2016} instead, the measured concentrations emerge as 16.8(\(\pm\)3.5) ppb and 12.0(\(\pm\)1.9) ppb, respectively, for the two sites. These values are well within the range of values previously reported for the region which lie between 6 ppb in Washington State and 156 ppb in Eastern Oregon \citep{doherty2014} and in summer snow on Mount Olympus, Washington State \citep{kaspari2015}. It appears that different reference values influence the accuracy of such a measurement for smaller concentrations. For concentrations \(>20\)ppb, this should not change the percent difference in inferred LAP concentration by nearly as much as in cleaner ice.
Although future work is needed to validate this estimate from laboratory analysis of ice from Collier Glacier, the range of values obtained demonstrates the utility of the measurements and the principle of the method. That said, laboratory analysis of LAP content and debris in snow and ice also suffers from limitations, with part of the particulate mass often not accounted for in dust and BC analysis. This leads to large uncertainty in the apparent optical properties estimated from laboratory analysis, further highlighting the utility of quantifying absorption independently from scattering. Both pathways - estimating LAP content from measurement of optical properties and estimating optical properties from LAP content - ultimately rely on precise knowledge of the mixing ratio and optical properties of each LAP. With the tight grouping of measured absorption coefficients shown in Figure \ref{fig:collier}, the technique presented here could provide reasonably accurate estimates of LAP concentrations in the field. This is particularly important for diagnosing cryospheric change in remote locations, where retrieval of ice cores and physical samples is logistically difficult and costly, and therefore presents a unique opportunity to simplify data collection. It is notable that bare ice albedo appears more sensitive to LAPs than snow, an effect that has been observed for sea ice as well \citep{marks2013}.
%This is consistent with the idea that impurities embedded within the glacier ice matrix, rather than externally mixed with air between snow crystals, may enhance total volumetric absorption. However, testing this hypothesis is particularly challenging for solid glacier ice relative to snow, which is easily excavated. The field method presented here therefore may open the door to testing this and similar questions of relevance to surface energy balance models and glacier melt rate estimates. (note - there is some literature on this idea about whether the LAPs are embedded within the ice crystal or are in the air cavities. you can imagine for snow how important it would be if the lap is within the crystal where all the path length is. I think Skiles is the right citation).

In summary, there are several valuable metrics that can be inferred from measured absorption and scattering coefficients. Since these measured values are effective properties of the bulk, no additional knowledge about the physical structure or assumptions about LAP content are needed, making these calculations particularly simple.

\section{Discussion}

We have shown that the portable instrument presented here can resolve photon time-of-flight (ToF) histograms in temperate, near-surface glacier ice. While this process works for short visible wavelengths at 405nm and 520nm, where analysis of the model fit performance revealed that the scattering and absorption coefficients are well constrained, this is not the case for the data measured at 640nm within the constraints of the temporal resolution of our experimental setup. Here, the ToF histograms are poorly resolved, which introduces significant correlations and poor accuracy in the retrieved parameters. Therefore, the data show taken at 640nm presented here is of limited quality and needs to interpreted as such. Improvements in temporal resolution would be desirable for red and infrared light. This concerns both the actual measurement resolution (currently 20ns) as well as the laser pulse duration of 18ns, which could be improved using commercially available means. For most data taken at 405nm and 520nm shown here, measured ToF histograms are of sufficient quality to extract optical scattering and absorption coefficients of bare glacier ice. While measurements taken at 640 nm are at the temporal resolution limit, impairing retrieval of the scattering coefficient, improved resolution could be enabled with a time-to-digital converter or FPGA-based electronics. The current setup can be implemented for roughly \(\$7,000\). While additional improvements such as a laser with higher pulse energy or shorter pulse duration would further reduce measurement duration and increase resolution and SNR, they would also increase cost. In addition to improvements in terms of resolution, validation of extracted LAP concentrations from laboratory measurements would be beneficial to evaluate the viability of using optical measurements as a proxy for LAP concentration. The same goes for bubble size distribution and concentration, which woulh help set bounds on the measured scattering coefficients. Another open question lies in the response of the measurement to inhomogeneities in the ice, which is not obvious from evaluating the current data set alone.

The coefficients measured here are near the upper end of the range of previously reported values in the literature \citep{cooper2020}, which implies that our setup can resolve ToF histograms on glaciers with similar composition.
For clean ice characteristic of polar regions, reported absorption coefficients of near-surface ice are smaller but, along with scattering coefficients, may exhibit greater spatial variability associated with greater range of ice histories exposed at the surface. For example, differences between the scattering and absorption coefficients of deep ice melting out near the terminus of large glaciers and ice sheets relative to younger ice near the equilibrium line altitude may be explainable in terms of a common age or ice provenance. Therefore, the ToF histogram approach presented here could provide additional context for optical coefficients measured by others in Greenland \citep{cooper2020} and Antarctica \citep{ackermann2006}, and could provide the basis for a systematic investigation of underlying structural controls on ice sheet albedo, which has been identified as a measurement priority needed to improve albedo parameterizations in large scale models \citep{dadic2013}.

Direct measurements of optical properties eliminate assumptions about structure and composition of ice. The simultaneous measurement of scattering and absorption properties allows us not only to infer LAP concentrations and albedo, but also to link these surface optical properties directly to the underlying features in composition and structure. With this complex set of parameters derived from a single measurement, it becomes possible to better understand the optical properties of bare ice and build useful albedo models. In addition, albedos inferred from scattering and absorption coefficients can be interpreted in terms of structure and composition, which is not practical with albedo measurements alone.
They in fact lie on the higher end of the range of previously reported absorption coefficients, which is expected for a mid-latitude glacier, where concentrations of LAPs can be expected to be higher than in the Arctic or Antarctic regions.
An example is the dependence of bare ice albedo on LAP concentration illustrated here: As the scattering coefficient is lower in bare ice as compared to snow, the dependence of the albedo on the absorption coefficient is stronger, making glacier ice albedo much more susceptible to LAPs. We observe significant albedo lowering of 0.07 from a difference in BC concentration of only 6ppb. In snow, such drastic lowering of the broadband albedo would require BC concentrations of several 1000ppb \citep{warren1980b}. In conclusion, albedo of higher density ice with lower bubble concentration like that found in the Greenland ablation zone \citep{cooper2020} would be highly susceptible to even small concentrations of LAPs.

Knowledge of the optical properties of glacier ice are needed to understand LiDAR interactions with snow and ice, including penetration depth, In addition, it has been shown that precise measurement of the directional dependence of the scattering coefficient can potentially identify the average crystal C-axis orientation and therefore flow direction  \citep{aartsen2013,katlein2014,rongen2020,rongen2021}. All this information can be made available in the field including rugged, remote terrain, reducing the need for sample shipping and off-site analysis, which could provide significant savings in cost and simplify expedition logistics. The non-invasive and small footprint nature of the instrument allows for deployment in remote areas such as the United States federally-designated Wilderness Areas, where motorized access and invasive measurement techniques are typically prohibited.

With more advanced techniques for photon-counting LiDAR such as photon-number threshold detection on the horizon \citep{cohen2019}, we hope to further improve the sensitivity to the point where it becomes possible to resolve the small scattering contributions of LAPs and dust. The individual behavior of contributions such as dust has only recently been measured \citep{cremonesi2020}, enabling us to potentially include such behavior into our model.

\section{Data availability} 

The raw data for all ToF histograms along with the code required to evaluate them and extract scattering and absorption coefficients are available at\\ https://doi.org/10.5281/zenodo.5828621.

\section{Code availability} 

The code used for evaluating each ToF measurement is published alongside the raw data at\\ https://doi.org/10.5281/zenodo.6262752. The LabView interface used to perform the measurement can be found at\\ https://doi.org/10.5281/zenodo.5828379. An online version of the COART code used to simulate the solar spectrum used in the broadband albedo calculation is available at\\ https://cloudsgate2.larc.nasa.gov/jin/coart.html.

\section{Acknowledgements} 
This work was supported by the University of Oregon through the Renee James Seed Grant Initiative. We thank Nicolas Bakken-French, Kyle Strachan and Jon Meyers for their help with field work.

\bibliography{main,related,theory,properties,aerosols,albedo,boreholes}
\bibliographystyle{igs}  % imposes IGS bibliography style on output

\appendix

\section{List of parts}

\paragraph{Lasers}
Thorlabs NPL41B, NPL52B, NPL64B
\paragraph{Detector}
Hamamatsu H7421-50 (most photon-counting PMTs can be used for this purpose)
\paragraph{Optical elements}
50mm focal length, 25mm diameter lens, anti-reflection coated for visible light: Thorlabs LA1131-A\\
10nm-bandpass filters: Thorlabs FB405-10, FB520-10, FB640-10\\
Lens tubes: Thorlabs SM1L10
\paragraph{Data Acquisition}
National Instruments USB-6361 with BNC breakout block (any NI USB-63xx series DAQ has the same counter functions necessary to run the LabView code).
\paragraph{Cables}
RG58 cables with BNC connectors\\
3dB attenuator for laser triggering
\paragraph{GPS Receiver}
Garmin GLO 2

\section{Time-of-flight measurements using NI data acquisition modules}

The NI DAQs have so-called "counters", which can either count edges of TTL (5V square) pulses on a specified input or act as pulse generators. To facilitate a time-of-flight measurements, we use three counters. \textit{Counter 0} generates a pulse train of predefined repetition rate and is supplied to an output, which is connected to the trigger input of the laser. Internally, the frequency output from \textit{Counter 0} is routed to the trigger input of \textit{Counter 1}, which also generates a pulse output, synchronized to \textit{Counter 0}. The output pulse length of the output of \textit{Counter 1} is equal to the desired gate length (time bin size). An artificial delay is applied between the trigger input from \textit{Counter 0} and the pulse output from \textit{Counter 1} which corresponds to the start time of the current time bin. The third counter, \textit{Counter 2}, works as an edge counter for the pulses coming from the photon-counting detector. The pulse train from \textit{Counter 1} is internally routed to the gate input on \textit{Counter 2}. To record the entire ToF histogram, the delay on the pulse output of \textit{Counter 1} needs to be step-wise increased.

Alternatively, if the laser can operate as a stand-alone pulsed source without external input, it can be used as a frequency master for the whole system. The laser needs an output, which provides a trigger signal whenever the laser fires. This trigger is then routed to an input on the DAQ. This input is then used to trigger a pulse train on \textit{Counter 0}, with a pulse duration equal to the desired gate length (time bin size). The Frequency output from \textit{Counter 0} is internally routed to the \textit{Gate Input} on \textit{Counter 1}, the TTL signal from the photon counting detector is routed to the input of \textit{Counter 1}. The artificial delay between the trigger input and the output of \textit{Counter 0} corresponds to the start time of the current time bin and needs to be increased step-wise to build a complete ToF histogram. With the lasers used here, this solution would work for repetition rates of 1MHz or faster, which the laser can produce without external trigger. In this configuration, a cheaper data acquisition module with only 2 counter circuits would suffice, but the setup would sacrifice the flexibility to use slower repetition rates which are required to record long ToF distributions as they would manifest for high-density, clean ice. This use of the DAQ is similar, but not identical to the "pulse separation" measurement described in the manual, which imposes further constraints on photon detection rates.

In both counter configurations, the repetition rate should be chosen so that the entire ToF distribution is captured and the measured intensity falls to the background count rate by the end of each cycle. The temporal bin size \(\Delta t\) can be chosen as small as two cycles of the counter circuit's clock frequency \(\nu_{\mathrm{clock}}\):

\begin{equation}
    \Delta t = 2 / \nu_{\mathrm{clock}}.
\end{equation}

\section{Approximate determination of the scattering-far-field "by eye"}

As a first approximation, the optical distance between source and detector can be determined "by eye", as the intensity falls off as a function of distance from the laser (see Eq. 33 in \cite{allgaier2021}):

\begin{equation}
    \phi(\rho) \propto \mathrm{exp}\left(-\sqrt{3\rho^2\sigma_{\mathrm{eff}}\sigma_{\mathrm{abs}}}\right).
\end{equation}

In ice, this exponential fall-off is dominated by the scattering coefficient, which is much larger than the absorption coefficient. A reduction in intensity of one order of magnitude corresponds to a distance of about 3 scattering lengths. As a simple approximation, the detector should be placed outside the area of visible backscattered light, ensuring several orders of magnitude in intensity reduction. However, if this distance is too large, there will be little backscattered light, reducing SNR. For small distances, arrival times will also be shorter according to Eq. \ref{eq:theo:bounded}, meaning that temporal resolution may not be sufficient to resolve the arrival time distribution appropriately.

%\section{Supplementary materials}

%We supply the following as supplementary material:
%\begin{itemize}
%    \item Raw data files
%    \item Matlab code for importing and analyzing raw histograms, including model fit functions
%    \item The LabView VI required for running the experiment
%\end{itemize}

\end{document}